\RequirePackage{fix-cm}
\documentclass[twocolumn]{svjour3}

\usepackage[english]{babel}
\usepackage[normalem]{ulem}
\usepackage{amsmath}
\usepackage{graphicx}
\usepackage{todonotes}

\usepackage{amsmath}
\usepackage{graphicx}
\usepackage{epstopdf}
\usepackage{lineno}

\usepackage{url}            

\usepackage{definitions}

\usetikzlibrary{shapes.geometric, arrows,calc,arrows.meta,positioning}
\tikzstyle{box} = [rectangle, rounded corners, minimum width=0.2cm, minimum height=1cm,text centered, draw=black]
\tikzstyle{arrow} = [thick,->,>=stealth]

\title{\boldmath LHC analysis-specific datasets with Generative Adversarial Networks}

\author{B. Hashemi \and
        N. Amin \and
        K. Datta\and
        D. Olivito \and
        M. Pierini
}

\institute{ B. Hashemi 
            \at Physics Department, University of California San Diego\\
            San Diego, USA
            \and
            N. Amin 
            \at Physics Department, University of California, Santa Barbara, USA 
            \and
            K. Datta
            \at Physics Department, ETH Z\"{u}rich, Z\"{u}rich, Switzerland
            \and
            D. Olivito
            \at Physics Department, University of California at San Diego, USA
            \and
            M. Pierini
            \at Experimental Physics Department, CERN, Geneva,\\ Switzerland
}

\begin{document}

\maketitle

\begin{abstract}
Using generative adversarial networks (GANs), we investigate the possibility of creating large amounts of analysis-specific simulated LHC events at limited computing cost. This kind of generative model is analysis specific in the sense that it directly generates the high-level features used in the last stage of a given physics analyses, learning the N-dimensional distribution of relevant features in the context of a specific analysis selection. We apply this idea to the generation of muon four-momenta in $Z\to \mu\mu$ events at the LHC. We highlight how use-case specific issues emerge when the distributions of the considered quantities exhibit particular features. We show how substantial performance improvements and convergence speed-up can be obtained by including regression terms in the loss function of the generator. We develop  an objective criterion to assess the geenrator performance in a quantitative way. With further development, a generalization of this approach could substantially reduce the needed amount of centrally produced fully simulated events in large particle physics experiments.
\end{abstract}

\section{Introduction}
\label{sec:intro}

In High Energy Physics (HEP), the use of Monte Carlo (MC) simulation techniques is a consolidated approach to characterize the experimental signature of a given signal hypothesis and to study potential background processes. This strategy was successfully applied many times in the last fifty years, e.g., to support the observation of neutral currents by Gargamelle~\cite{Gargamelle} or, more recently, to optimize the search for the Higgs boson at the CERN Large Hadron Collider (LHC)~\cite{HiggsATLAS,HiggsCMS}. The time distance between these two publications gives an indication of how consolidated this practice is in HEP.

The setup of a typical HEP experiment foresees sufficient computational resources to provide analysts with large sets of MC simulated events. Though it has not been the case historically, it is becoming common for LHC physics results to suffer from limited statistical precision in their MC datasets, due to the high CPU and size-on-disk cost of generating and storing MC events (see for instance Refs.~\cite{Sirunyan:2017uzs,CMS-PAS-TOP-18-006}). Currently, accurate detector response simulations, typically based on the {\tt GEANT4}~\cite{GEANT4} library, are among the most CPU-intense workflows on the  LHC computing grid, as discussed, for instance, in Ref.~\cite{Hernandez:2008zzb} for the CMS experiment.

With the perspective of integrating as much as 300 fb$^{-1}$ by the end of Run III and as much as 3000~fb$^{-1}$ during the High-Luminosity LHC (HL-LHC) phase~\cite{Apollinari:2017cqg}, the LHC experiments (particularly ATLAS and CMS) will face a substantial increase of computing needs for MC event processing. In general, there will be a need for larger MC samples, to match the larger collected dataset. Moreover, event simulation and reconstruction processes will become much more computationally intense, with more granular detector components to be modelled and as many as 200 collisions overlapping with the event of interest (pileup). 
The increase in required resources will not be matched by an equivalent increase in computational power, as discussed for instance in Ref.~\cite{CMS_TP} for the CMS experiment. 
This limitation would offset the advantage coming from the large collected dataset and put under question the need to go at such a high luminosity. This is why the HEP community is investigating possible directions to speed up the detector simulation~\cite{GEANTV,ManyCoreGEANT4}.

Generative Adversarial Networks (GANs)~\cite{GAN} offer a concrete possibility to speed up MC generation, as already demonstrated in recent studies (see Sec.~\ref{sec:relatedWork}). These efforts mostly attempt to replace specific steps of the event generation+reconstruction pipeline with generative models, resulting in a substantial reduction of CPU usage. On the other hand, the needed CPU resources might still be substantial. For instance, generating single-particle showers and jet images with GANs~\cite{JetGAN,CMSOpenJet} one would still have to perform tasks such as particle tracking, event selection, and computation of high-level quantities used in the final signal-extraction statistical analysis. This point is addressed in Ref.~\cite{JetGAN} in the context of jet simulation. There, a GAN model is trained to return reconstructed jets as images. However, this is not necessarily a general solution. For instance, an event-as-image representation might not perfectly fit a reconstruction based on Particle Flow~\cite{particleflow}.

A typical reconstructed MC event consists of $\sim 1$~MB of data. However, a typical analysis would only use $\sim 10$~kB of high-level features, computed from reconstructed particles in the event. If one could directly generate the information pertaining to the relevant set of particles in the event, i.e. the analysis-specific $\sim 10$~kB of data, the total CPU and size-on-disk savings would be measured in orders of magnitude. In this study, we attempt to short-circuit the entire event generation process, using a GAN model to go straight from random numbers to a complete reconstructed event, represented not in terms of raw detector hits but as a vector of quantities needed in the last stages of a specific analysis. This strategy would come with tremendous CPU and storage saving, since the final $\sim 10$~kB of high-level features could be directly produced, bypassing any intermediate processing and reducing needs for sample storage.

In this paper, we implement a proof-of-concept for generating analysis-specific datasets with GANs. As our example, we consider a hypothetical analysis which uses $\cal{O}$(20) features from a dimuon final state, e.g. lepton momentum vectors, isolation, and jet transverse momenta. We use GANs to learn the multi-dimensional distribution of a Drell-Yan sample and generate new events following the same distribution. This would be similar to using numerical representations of a generic function, e.g. with kernel methods. On the other hand, the use of neural networks offers in general better scaling performance with the number of dimensions.

For a considered dataset of $\cal{O}$(10) quantities, a generator and discriminator with $\cal{O}$(200,000) tunable parameters spread over roughly 10 layers provides reproducible and satisfactory performance. We obtain good accuracy on a minimal setup and highlight use-case-specific issues that degrade precision whenever the distribution of the quantities of interest exhibit edges at the boundary of the definition range, have multiple peaks, or are discrete in nature. Use-case-specific workarounds are discussed. This paper is intended to demonstrate the potential of this strategy, but its outcome cannot be taken as a conclusive and ready-to-use algorithm. A more robust R\&D program will need to be undertaken to consolidate this approach beyond the small-scale demonstration presented here.

Nevertheless, the advantages of the proposed strategies are already clear. The speed-up factor for the full generation process is found to be substantial, even when compared to the fast-simulation workflow used to generate our reference events. In particular, our final generative model encodes a 2 GB dataset while taking up less than 10 MBs on disk, and is capable of producing over 5000 events per second. For comparison, generating 5000 events (with {\tt PYTHIA}~\cite{pythia}) and reconstructed them (with {\tt DELPHES}~\cite{delphes}) requires $\sim 2$ and $\sim 1.5$ minutes on a 3~GHz Intel i5, respectively. A GEANT-based detector simulation and a full event reconstruction (including tracking) would be typically two orders of magnitude slower. 

Besides presenting a practical implementation of our idea, we propose a best-model selection strategy, based on a quantitative assessment of  the generation accuracy. Such a procedure allows to deal with the known instabilities of GAN training and to avoid the kind of by-eye quality assessment which is often adopted for GAN applications.


This paper is organized as follow: related works are briefly discussed in Section~\ref{sec:relatedWork}. Section~\ref{sec:data} describes the dataset and features used. The network models and training procedure are presented in Section~\ref{sec:model}, and the results are shown in Section~\ref{sec:Results}. In Section~\ref{sec:remarks}, we discuss the quality of the results and address possible limitations in generalizing this approach to different datasets. Conclusions are given in Section~\ref{sec:conclusions}.

\begin{figure*}[tb!]
\centering
\includegraphics[width=\textwidth]{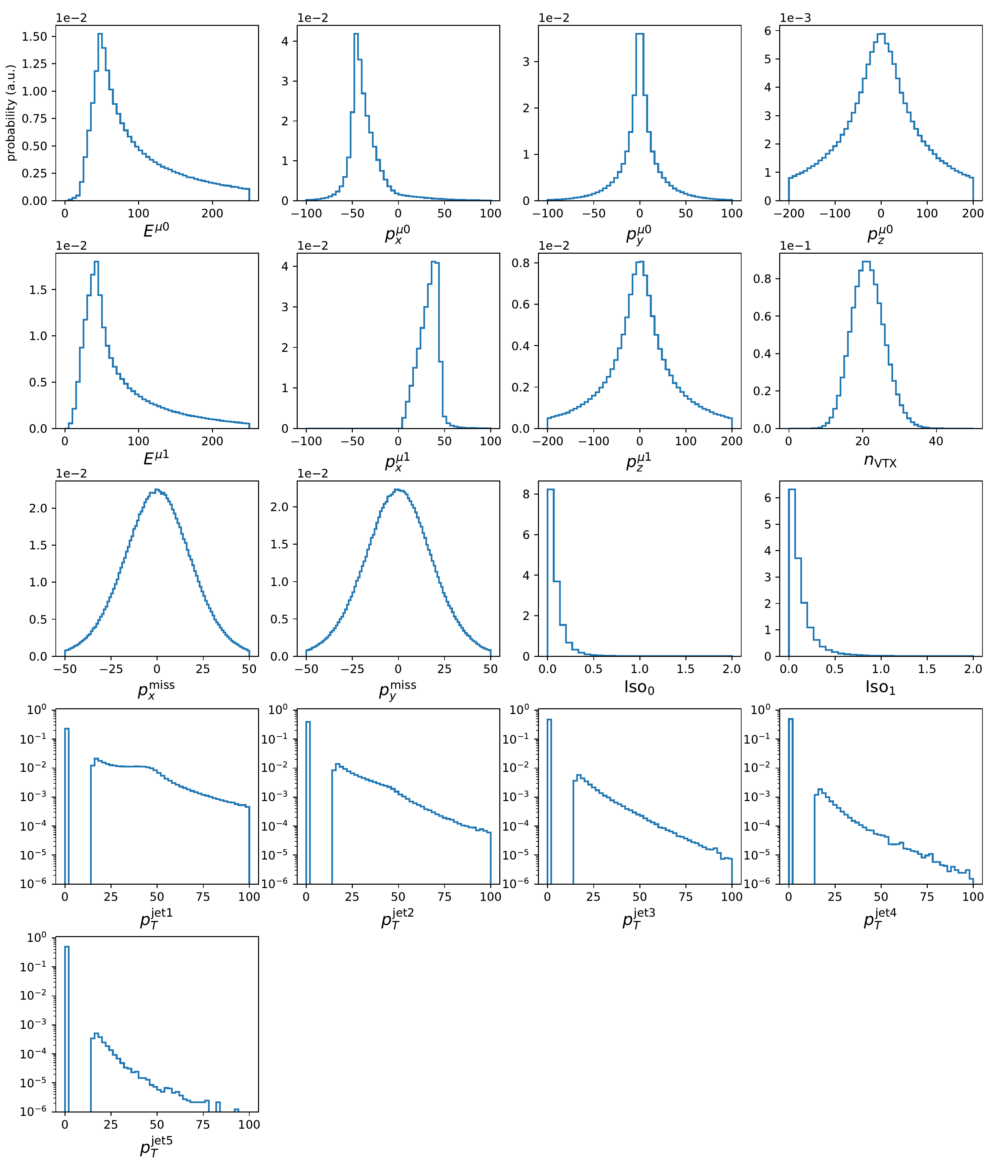}
\caption{Distribution of the selected features in the target dataset, Drell-Yan to dimuon events generated with {\tt PYTHIA8} and reconstructed with the CMS detector simulation in {\tt DELPHES}  \label{fig:inputs}}
\end{figure*}

\section{Related Work}
\label{sec:relatedWork}

Generative adversarial networks~\cite{GAN} have been investigated for LHC applications to simulate the energy deposits of individual particles~\cite{CaloGAN1,CaloGAN2,LcdGAN} and jets~\cite{JetGAN,CMSOpenJet}, as well as to accelerate Matrix-Element methods~\cite{matrix_ml}. Recently, a GAN-based generator was developed to simulate data collected in test-beam studies for the future CMS Highly Granular Calorimeter~\cite{Erdmann:2018jxd}. A similar study was carried on in the context of cosmic-ray detection~\cite{Erdmann:2017str}. A discussion of how GAN models could be relevant to event simulation in future HEP experiments can be found in Ref.~\cite{Apostolakis:2018ieg}. 

Typically, GAN applications in HEP focus on image representations of the collected data. Instead, this work is based directly on high-level features computed from the reconstructed events. To our knowledge, the only other work exploring this possibility is Ref.~\cite{Otten:2019hhl}, which appeared while we were finalizing this paper. Ref.~\cite{Otten:2019hhl} describes the same idea proposed here, with similar methodology and results. In addition, improved performances are obtained when using variational autoencoders, previously investigated in Ref.~\cite{Cerri:2018anq} in the context of anomaly detection. 

The adversarial training (AT) technique is  used in HEP for tasks other than event generation: reference~\cite{Louppe:2016ylz} discusses how to account for uncertainties associated to a given nuisance parameter using AT. Reference~\cite{Shimmin:2017mfk} uses AT to preserve the independence of a given network score (a jet tagger) from a specific physics quantity (the jet mass). This technique was also used to train autoencoders for jet tagging~\cite{Heimel:2018mkt}. Reference~\cite{Datta:2018mwd} discusses how to use a GAN setup to unfold detector effects. 

\begin{figure*}[tb!]
\centering
\includegraphics[width=0.75\textwidth]{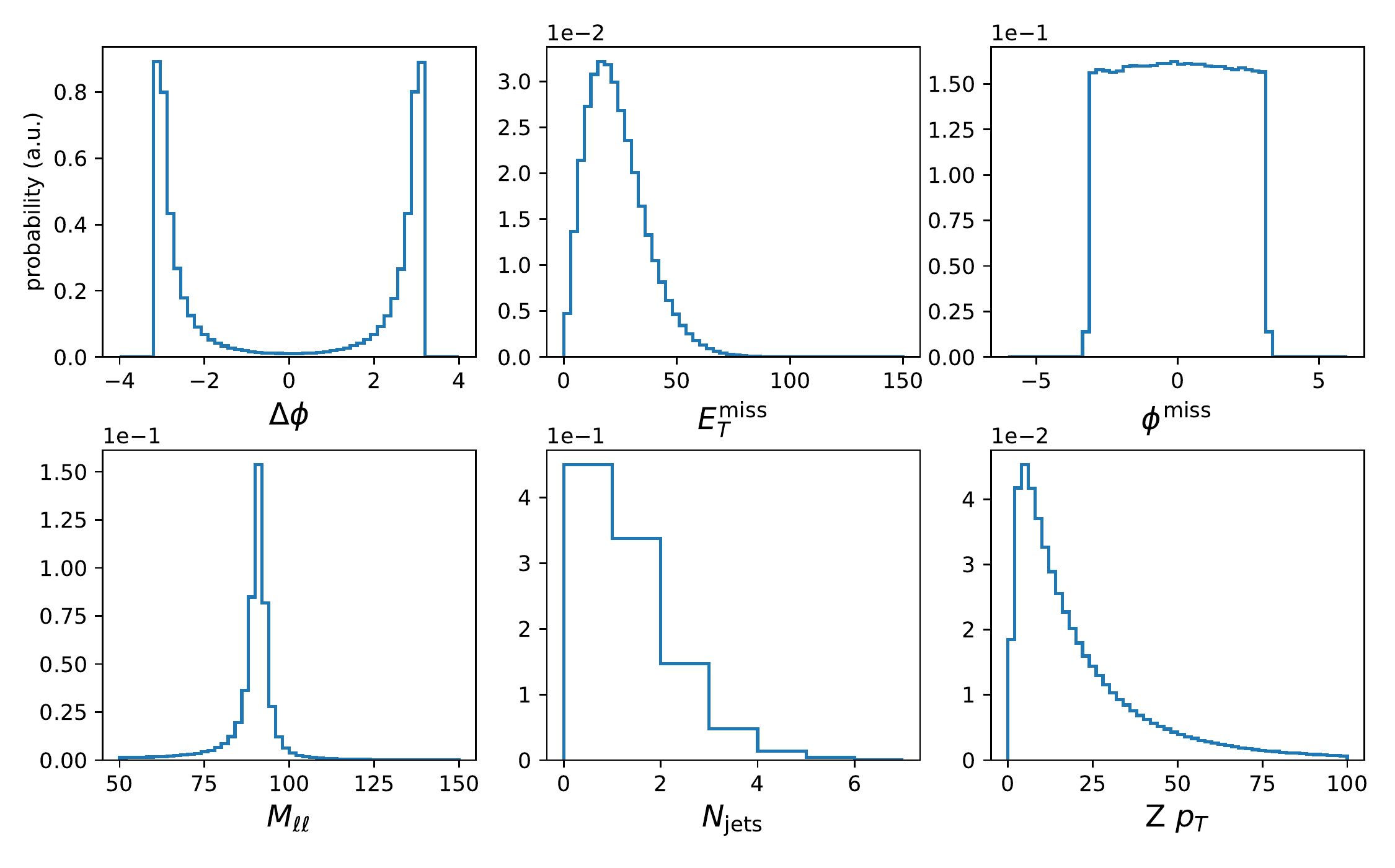}
\caption{Distribution of six quantities computed from the 18 considered features of the $Z\to \mu\mu$ dataset.\label{fig:derived}}
\end{figure*}

\section{Dataset description}
\label{sec:data}

We consider a sample of $Z \to \mu \mu$ events in proton-proton collisions, generated using the {\tt PYTHIA8}~\cite{pythia} event generator at a center-of-mass energy of 13~TeV. Detector resolution and efficiency are taken into account using the parametric description of the upgraded CMS detector for High-Luminosity LHC~\cite{Chatrchyan:2008aa,CMS_TP} provided by the {\tt DELPHES} detector simulation library~\cite{delphes}. Events are generated with an average of 20 simultaneous collisions (pileup), similarly to what the LHC delivered in 2015-2016. 

Events are represented as an array of numbers, corresponding to the following 17 features:
\begin{itemize}
\item The four-momenta ($p_{x}^{\mu i}$, $p_{y}^{\mu i}$, $p_{y}^{\mu i}$, $E^{\mu i}$) of the two muons ($i=0,1$) in Cartesian coordinates.\footnote{As is common for collider physics, we use a Cartesian coordinate system with the $z$ axis oriented along the beam axis, the $x$ axis on the horizontal plane, and the $y$ axis oriented upward. The $x$ and $y$ axes define the transverse plane, while the $z$ axis identifies the longitudinal direction. The azimuth angle $\phi$ is computed from the $x$ axis. The polar angle $\theta$ is used to compute the pseudorapidity $\eta = -\log(\tan(\theta/2))$. We fix units such that $c=\hbar=1$ and give all energy-based quantities in units of GeV.} In early stages of this work, a cylindrical-coordinate representation of the muon four-momenta was considered, which resulted in a degradation of the generator performances.
\item The number of reconstructed primary vertices in the event, $n_{\mathrm{VTX}}$.
\item The Cartesian coordinates ($p_x^{\mathrm{~miss}}$, $p_y^{\mathrm{~miss}}$) of the missing transverse momentum $\vec{p}_T^{\mathrm{~miss}}$, defined as the negative vector sum of transverse momenta for all reconstructed particles in the event. Its absolute value, the missing transverse energy (\MET), provides a measurement of the transverse momentum carried by undetected particles. 
\item The muon isolation, computed as the sum of the transverse momenta in a cone around the muon, normalized by the muon transverse momenta, from three classes of particles: charged particles, photons, or neutral hadrons. For each particle class, the precise definition for the isolation is:
\begin{equation}
{\rm Iso} = \frac{\sum \limits_{i \neq \mu} p_T^i}{p_T^{\mu}}~,
\end{equation}
where the index $i$ refers to particles of the appropriate class, excluding the muon itself in the case of charged objects, within angular distance $\Delta R = \sqrt{(\Delta \eta)^2+(\Delta \phi)^2} < 0.3$ from the muon.
\item The transverse momenta of the five highest-$p_T$ jets in the event, clustered, using the anti-$k_{T}$ algorithm with a radius parameter $R=0.4$ \cite{antikt}. If less than five jets are found,  \pt=0 is assigned to additional jets.
\end{itemize}

With no loss of generality, all features related to the two muons are ordered such that $i=0$ ($i=1$) corresponds to the muon with highest (lowest) \pt. 

In order to facilitate the generator training, we pre-process the inputs as follows: 
A rotation of the two four-momenta is applied, so that $p_y^{\mu 1}=0$, after the rotation. Once this is done, $p_y^{\mu 1}$ is discarded from the dataset. 
Then, the number of vertices, \nvtx, a discrete quantity, is smoothed by replacing the integer value with a floating point number sampled from a Gaussian centered at the integer value with width $\sigma=0.5$. For instance, an event with 25 vertices is assigned an $n_{\mathrm{VTX}}$ value sampled from the probability density function (pdf) $\frac{1}{\sqrt{\pi}}e^{-2(x-25)^2}$. When the generator is used to create new events, we take the floor of \nvtx, making it a discrete quantity again. The processed distributions of these 17 features are shown in Fig.~\ref{fig:inputs} for the target dataset~\footnote{The data used for this project can be downloaded from the link:
\url{http://uaf-10.t2.ucsd.edu/~bhashemi/GAN_Share/total_Zmumu_13TeV_PU20_v2.npa}.}.

For the purpose of verifying how well the generator predicts correlations between quantities, we also inspect the following quantities, computed from the 17 input features: 
\begin{itemize}
\item The dimuon invariant mass, $\mll$.
\item The transverse momentum of the dilepton system, corresponding to the \pt of the $Z\to\mu\mu$ boson. 
\item The event \met and its azimuth angle \metphi.
\item The angular separation of the two muons in the transverse plane, $\Delta\phi$, defined in the interval $[-\pi,\pi]$.
\item The jet multiplicity, $N_\mathrm{jets}$, computed by counting the number of jets with $\pt>15$ GeV.
\end{itemize} 
The distributions of these six quantities are shown in Fig.~\ref{fig:derived}.

\section{Model definition}
\label{sec:model}
A detailed explanation of how to build and train GAN models can be found in the original paper~\cite{GAN} and in many of the application papers discussed in Section~\ref{sec:relatedWork}. In this section, we introduce two implementations of the application presented in this work, designed for two specific use cases. In the first, the Drell-Yan dataset of Section~\ref{sec:data} is reduced to the two muon four-momenta. In the second, the full dataset is considered. 

\begin{figure*}[t]
\centering
\includegraphics[width=\textwidth]{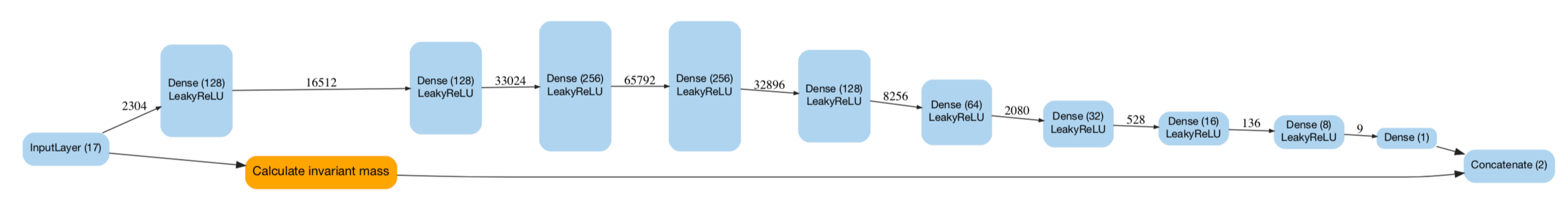} \\
\includegraphics[width=\textwidth]{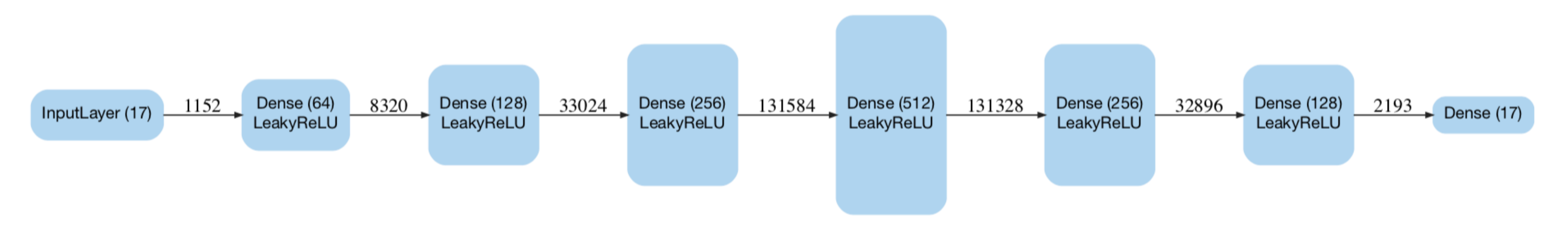}
\caption{Network architectures for the discriminator (top) and generator (bottom).\label{fig:models}}
\end{figure*}


The same network architecture is used for both cases (see Fig.~\ref{fig:models}), besides adapting the length of the input and output layers of the generator and the input layer of the discriminator to fit the dataset dimensionality $N$ of each problem (17 for the full dataset and 7 for the reduced one). The generator network consists of 7 fully connected layers with 64, 128, 256, 512, 256, 128, $N$ neurons. Neurons in the inner layers are activated by leaky ReLU functions, while linear activation functions are used for the output layers. The input to the generator network consists of $N$ ``noise'' floating-point numbers, sampled from a Gaussian distribution centered at 0 with unit variance.  In early stages of this work, we  tried to increase the model size, by increasing the number of input noise variables beyond $N$ and using wider inner layers. These changes caused frequent problems of mode collapse before the training could reach equilibrium, a frequent issue when training GANs. The final architecture of Fig.~\ref{fig:models} was chosen to limit this problem.

The discriminator network consists of 9 hidden dense layers with 128, 128, 256, 256, 128, 64, 32, 16, and 8 neurons, activated by a leaky ReLU function. The last hidden layer is fully connected to a single-neuron output layer with sigmoid activation. In addition, a layer connected directly to the input layer returns the dilepton mass as part of the output.
The per-batch distribution of this quantity is used to the loss function. This additional ingredient allows to stabilize the training and improve results, as discussed below.

The combined network is trained adversarially for 100,000 single-batch epochs for the full model with $N=17$, and 40,000 epochs for the simplified model with $N=7$. In both cases, the batch size is fixed to 512 events. Each batch consists of random examples from the full dataset (1.6 million examples). 

The network training proceeds as follows: first, the discriminator network is given a half-batch of 256 examples of real events and 256 examples of generator output. In a pre-training stage, a binary cross entropy (BCE) loss function is minimized: the network is trained to output 1 for the real examples and 0 for the examples returned by the generator. Then, the generator is trained using a three-term loss function:

\begin{equation}
\label{eq:gen_loss}
{\rm L} = \text{BCE} + c_1 \cdot \left( \mll - M_Z \right)^2 + c_2 \cdot \left( \sigma_{\mll} - \sigma_{M_Z} \right)^2 .
\end{equation} 

In Eq.~(\ref{eq:gen_loss}), the second term is the square of the average deviation from the mass of the Z boson over all generated events in the batch. The final term is the squared difference between the standard deviation of the \mll values for the batch produced by the generator, and the standard deviation of the \mll values in the target dataset. We fix $M_Z=89.6$~GeV and $\sigma_{M_Z} = 7.7$~GeV, corresponding to the mean and RMS of the $\mll$ distribution in the input dataset. The $c_1$ and $c_2$ terms in Eq.~(\ref{eq:gen_loss})  guarantee that the three contributions to ${\rm L}$ are of comparable size. Training our models, we verified that changing $c_1$ and $c_2$ has little impact on performance. Because of this, we simplify the loss function forcing $c_1=c_2=0.0001$ with no loss of generality. 

\section{Results}
\label{sec:Results}

All networks were implemented in {\tt KERAS}~\cite{keras}, using {\tt TensorFlow}~\cite{tensorflow2015-whitepaper} as a back-end. The code is available on GitHub~\cite{codebase}. The training was carried on using the adadelta optimizer~\cite{adadelta}. We also ran experiments using the adam~\cite{adam} optimizer, for which a deterioration of the generator performances was observed. 

\begin{figure*}[t]
\centering
\includegraphics[width=\textwidth]{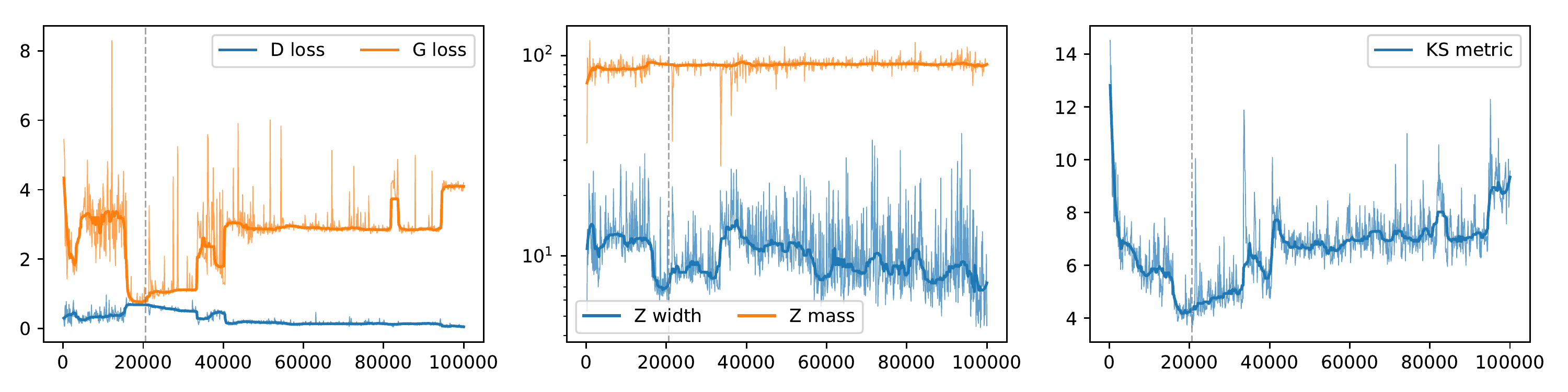}
\includegraphics[width=\textwidth]{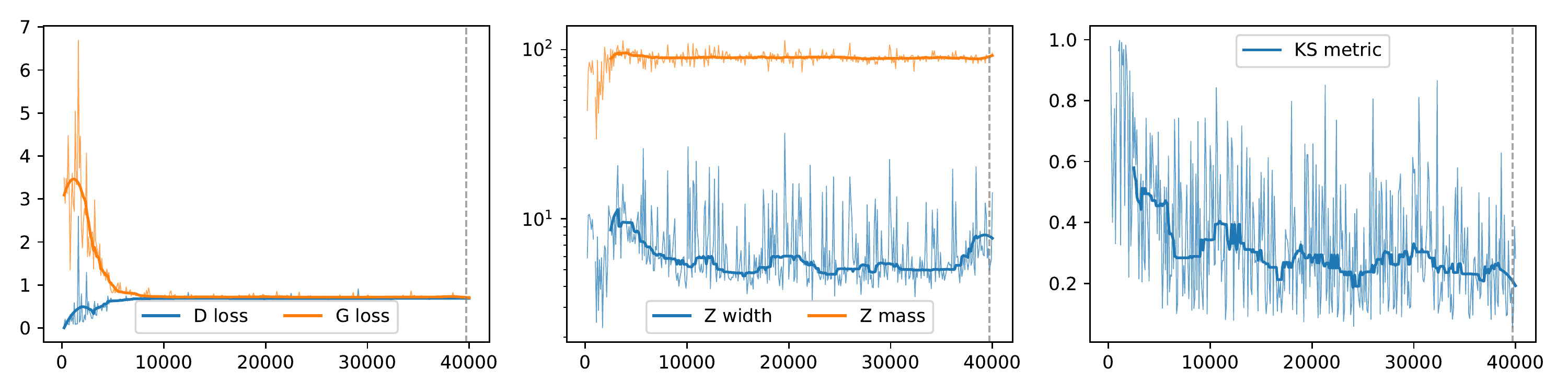}
\caption{Training history for a typical network training on the reduced (top) and full (bottom) dataset. Left: Discriminator and generator loss as a function of the training epoch. Center: Mean and standard deviation of invariant masses calculated from generated samples. Right: Sum of Kolmogorov-Smirnov test statistics across all marginal distributions. \label{fig:training}}
\end{figure*}

The training was performed using an NVIDIA GeForce GTX Titan X. In addition, part of the work was carried on at the Piz Daint supercomputer of the Swiss National Supercomputing Center (CSCS), mounting NVIDIA Tesla P100 GPU cards. To achieve the best performance, the training was repeated 100 times for the full model with $N=17$ features, and 10 times for the reduced model with $N=7$. The evolution of the loss function in Eq.~(\ref{eq:gen_loss}) as a function of training epoch is shown in Fig.~\ref{fig:training} for the trails whose results are displayed in this section.

 To assess training convergence, network weight values are saved every 100 epochs. For each of these saved training snapshots, a Kolmogorov-Smirnov (KS) test is performed on $\mll-M_Z$, \metphi, and ${\rm Iso}_1$, comparing the distribution of these features in the original dataset to what is produced by the generator. Using the sum of the three KS scores as a figure of merit, the collected training snapshots are assessed, as shown in  the right plot of Fig.~\ref{fig:training} for one of the large-model trainings.
For each training, the outputs of the top-20 networks were further inspected by eye, to select the best performing network. The networks with the lowest summed rank were found to give very similar performances. In addition, the best KS scores are found to occur in coincidence with an overall good description of all the features in the dataset. This allows to reliably assess the overall generator performance by looking at only a subset of the generated quantities. On the other hand, the procedure could be easily extended to the full set of features, if required by specific aspects of a given input dataset.

\subsection{Training on Reduced Dataset}

\begin{figure*}[t]
\centering
\includegraphics[width=\textwidth]{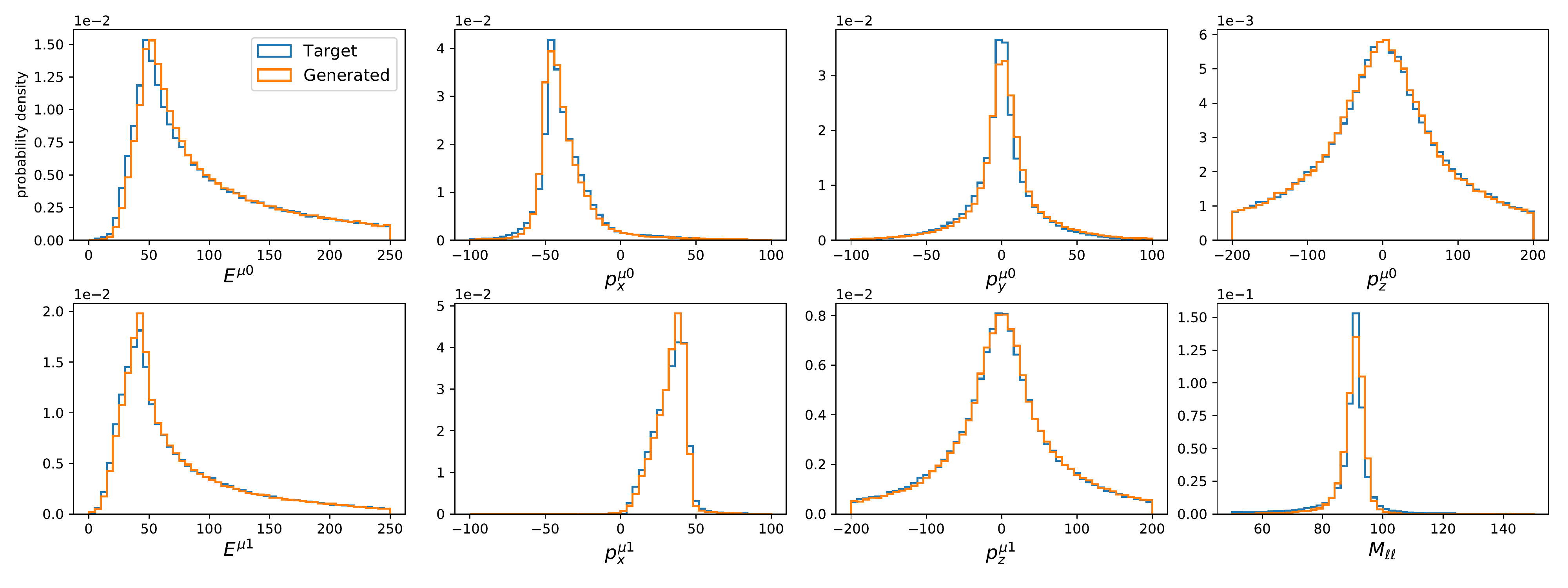}
\caption{Comparison between the distribution of input and GAN-generated quantities, and for the $\mll$ distribution derived from them and entering the minimized loss function defined in Eq.~(\ref{eq:gen_loss}). \label{fig:results_minigan}}
\end{figure*}

Figure~\ref{fig:results_minigan} shows the distributions returned by the generator for the inputs of the reduced model (two muon four-momenta and the dilepton mass distribution), for one of training snapshots with the top-20 KS score sums. An overall good agreement is observed. 

\begin{figure*}[t]
\centering
\includegraphics[width=\textwidth]{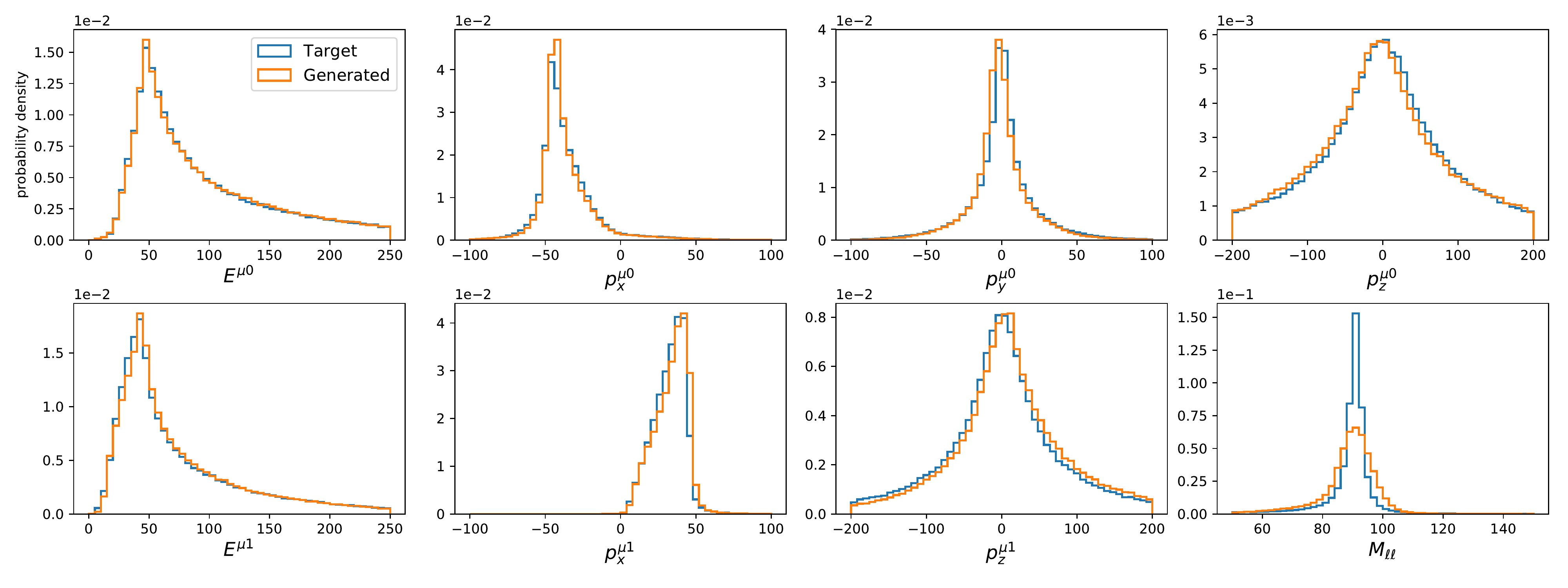}
\caption{Comparison between the distribution of input and GAN-generated quantities, and for the $\mll$ distribution derived from them, obtained removing the two $\mll$-related terms from the loss function in Eq.~(\ref{eq:gen_loss}).\label{fig:results_minigan_bce}}
\end{figure*}

Figure~\ref{fig:results_minigan_bce} shows a similar result, obtained setting $c_1=c_2=0$ in Eq.~(\ref{eq:gen_loss}). The four-momenta of the two muons are still well reproduced, with some performance loss observed (e.g., in $p_z^{\mu 2}$). 
As expected, once any relation between the $\mll$ distribution and the loss is removed, a strong performance degradation is observed in the description of $\mll$.

The comparison of this result to those of Fig.~\ref{fig:results_minigan} highlights how the network architecture should be adapted to the specific dataset one wants to generate: post-processing should be foreseen for each interesting function of the input quantities, and a set of corresponding terms should be added to the loss function to force the learning of the quantity's distribution. Once the network is forced to learn these extra quantities, 
it is possible to achieve reliable analysis-specific dataset generation. This  provides a viable alternative to methods such as density kernel estimation for approximating pdfs on arbitrary data, with the advantage of offering a better scalability with increasing  dataset dimension.

\subsection{Training on Full Dataset}

\begin{figure*}[htb!]
\centering
\includegraphics[width=\textwidth]{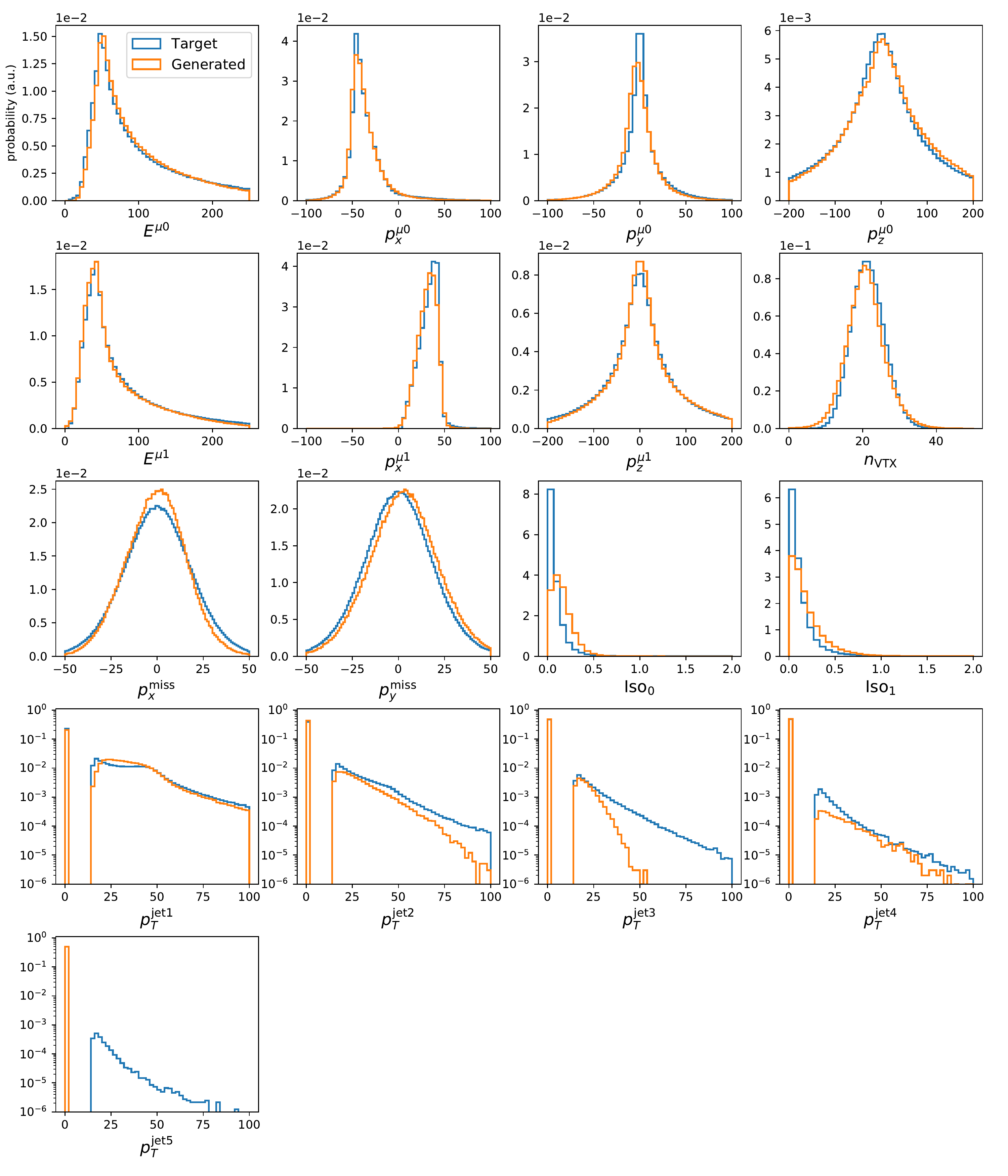}
\caption{Comparison of the target distributions to those generated by the generator model for the full dataset.\label{fig:results_fullgan}}
\end{figure*}

\begin{figure*}[htb!]
\centering
\includegraphics[width=0.75\textwidth]{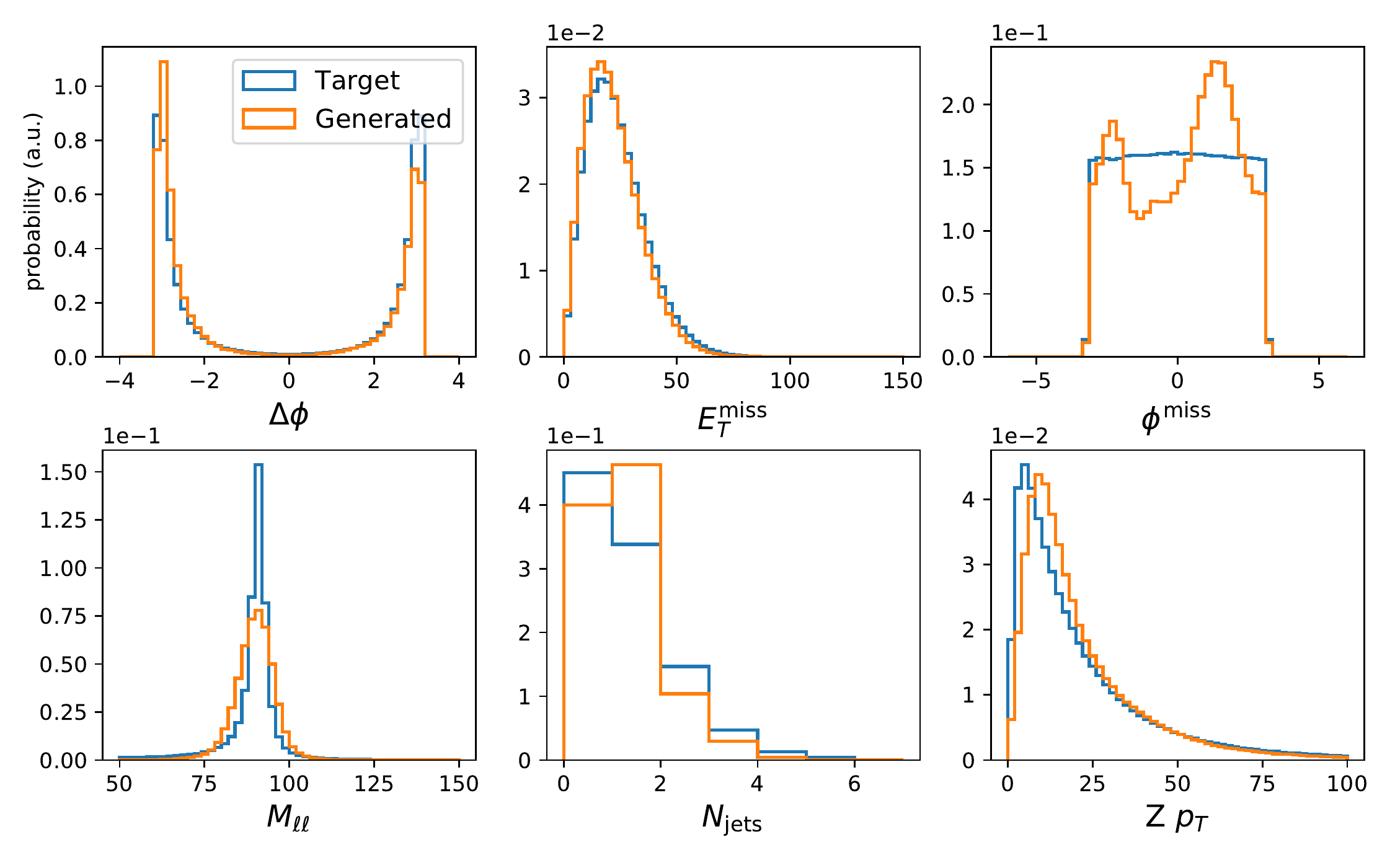}
\caption{Comparison target and input distributions for a set of quantities derived from the 17 quantities returned by the GAN.
\label{fig:results_fullgan_derived}}
\end{figure*}

Following the training quality assessment described in the previous section, we trained a GAN model on the full dataset. Figure ~\ref{fig:results_fullgan} shows a comparison between the target and generated distributions for all 17 input quantities defining the training dataset, where the generator model was selected as the best performer from the top-5 models selected by the KS sum test. Figure~\ref{fig:results_fullgan_derived} additionally shows a comparison for meaningful physics quantities derived from the 17 features generated by the same generator model. 

The GAN model is still capable of describing the muon four-momenta with accuracy, at least in the marginal distributions, but less perfect agreement is observed for the lepton isolation variables and the jet $\pt$. Both quantities are characterized by a sharp edge at the lower boundary of the range in which these features are defined. Such boundaries and sharp features appear to be difficult to learn for the GAN, 

In this respect, we found it more convenient to use, when possible, Cartesian coordinates to represent four-momenta. A further complication with the jet $\pt$ distribution comes from the presence of a spike at zero, disjoint from the rest of the distribution. This is an artifact produced by zero-padding for events having small jet multiplicity. This problem could be circumvented training different models on events with different jet multiplicities. This is just an example of the many case-specific workarounds that could be played to simplify the GAN's task (see Appendix~\ref{sec:discrete_quantities} for a similar discussion on discrete quantities). On the other hand, we found interesting to discuss the result of a generic training, to show not just the success but also any encountered limitation, to offer a sense of where the consolidation effort should go in the future.

It is also interesting to notice that the symmetry in $\metphi$ is broken, due to small deviations in the transverse plane's missing momentum. This demonstrates the importance of choosing input variables that are most important for the analysis. Additionally, as a consequence of the imperfect modelling of the jet $\pt$s, the jet multiplicity deviates from the target dataset to a moderate degree (see Fig.~\ref{fig:results_minigan_bce}). 

Figure \ref{fig:training} shows that the full model does not seem to be able to converge to a stationary minimum in the loss function. This is likely why perfect agreement in the \mll distribution is not achieved with the full model, even with the explicit dependence of the loss function on the mean value and width. Even so, reasonable performance can be achieved by cherry-picking an epoch where the cost function is near a minimum with the help of the additional KS based metric described in this section.

\begin{figure*}[htb!]
\centering
\includegraphics[width=\textwidth]{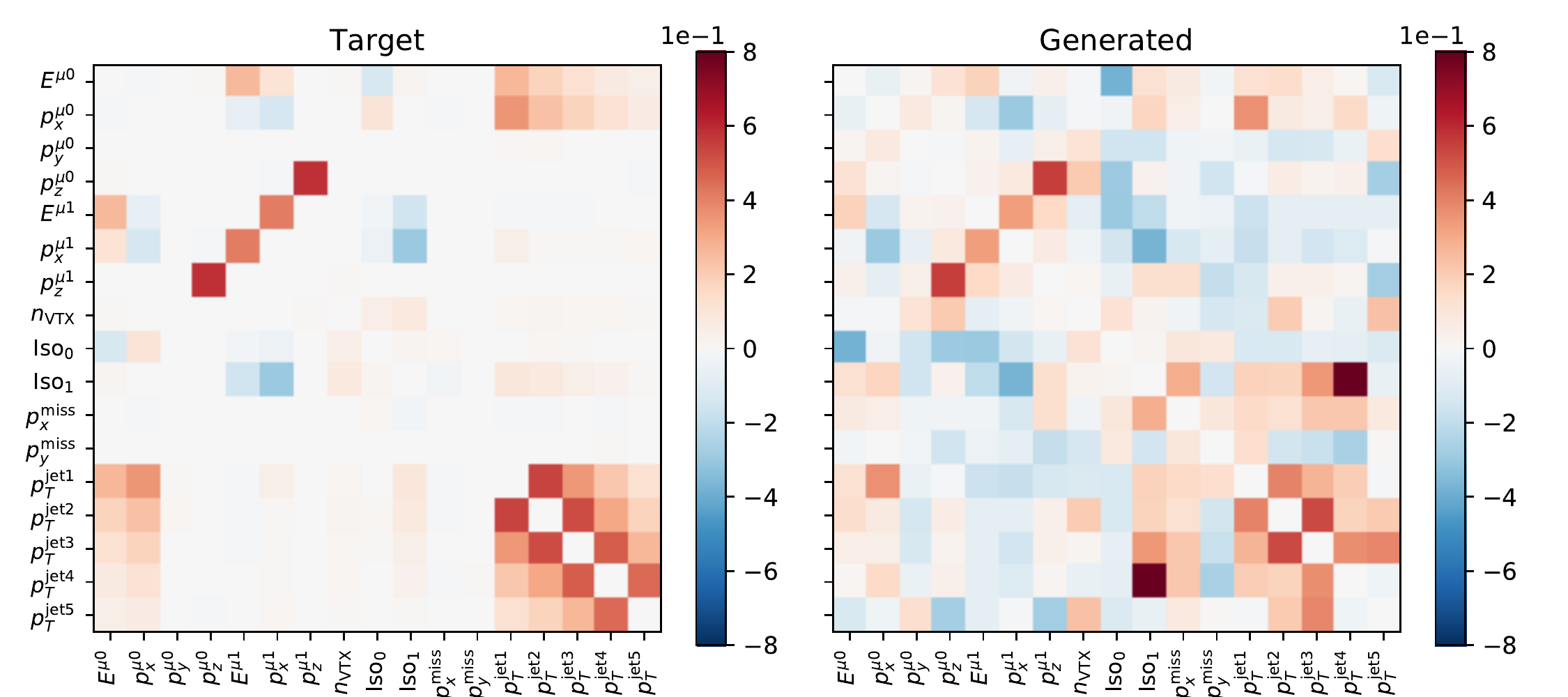}
\caption{Pairwise correlation matrices of features for target (left) and generated (right) samples, the diagonal terms have been set to 0 for aesthetics. The pdf for this model is not expected to be a high dimensional Gaussian, so the correlation matrix must be taken with a grain of salt.\label{fig:comparisonCorrelation}}
\end{figure*}

\begin{figure*}[t]
\centering
\includegraphics[width=.48\textwidth]{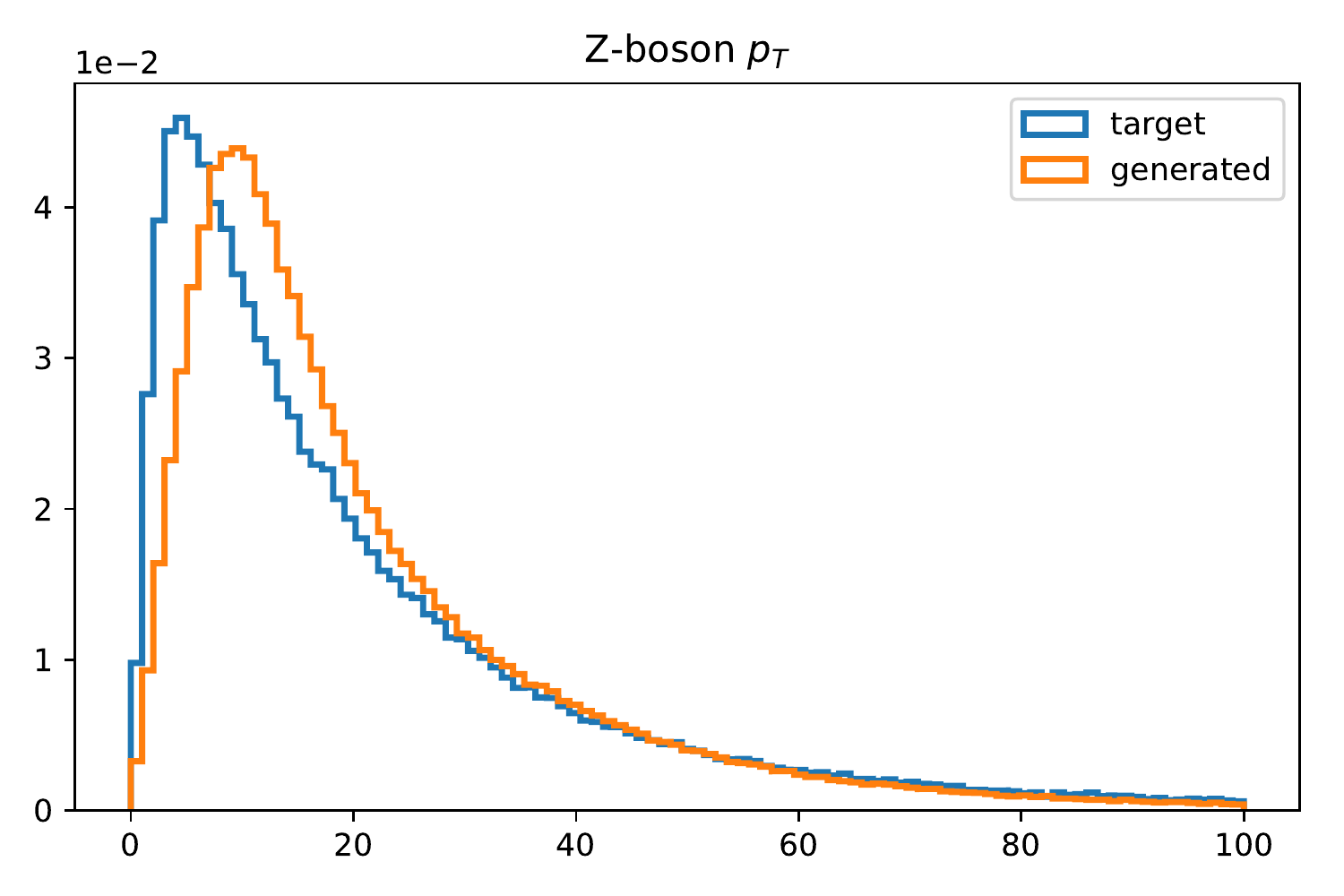}
\includegraphics[width=.48\textwidth]{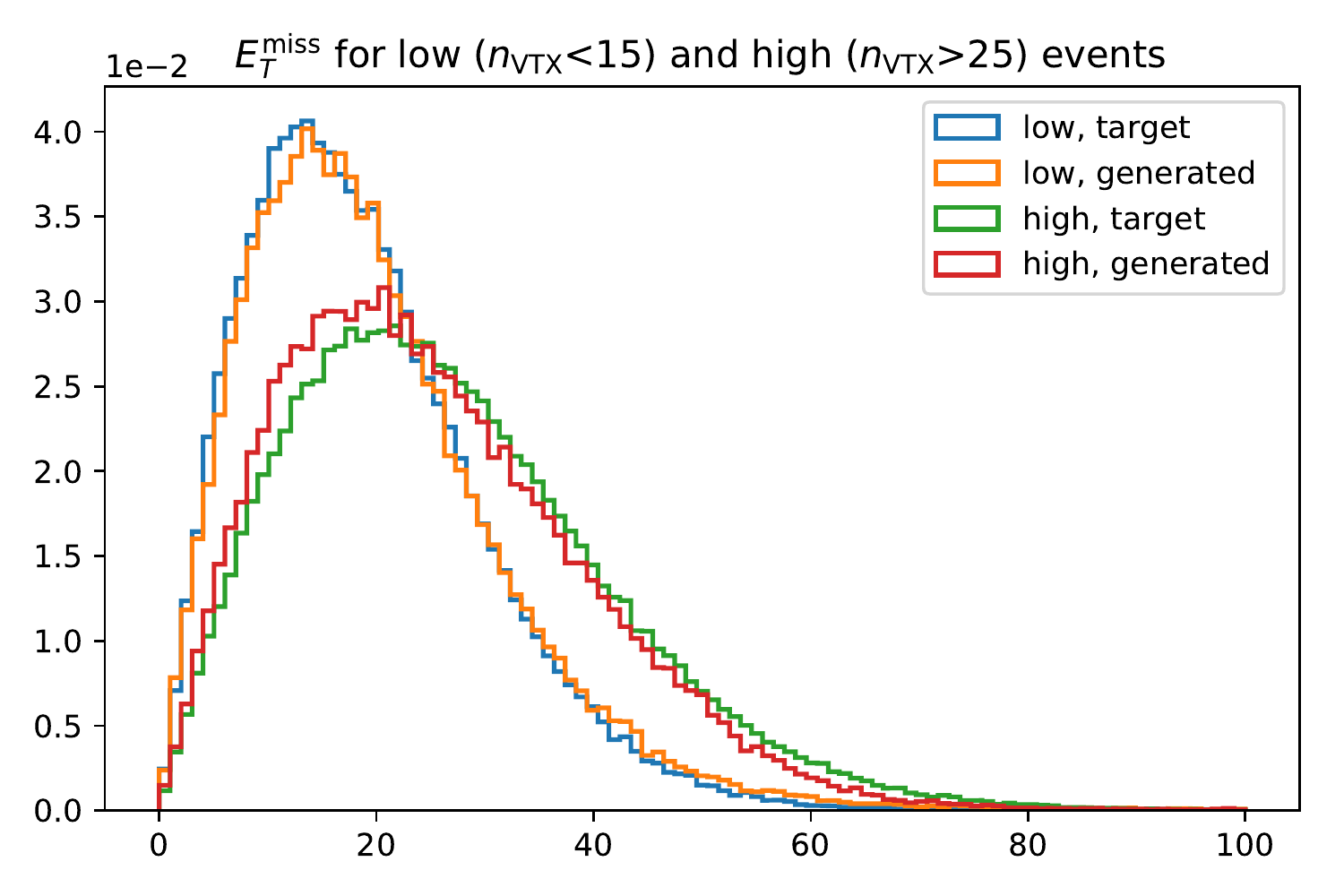}
\caption{Left: Z-boson transverse momentum in the target and generated datasets. Right: Comparison of the $E_T^{\mathrm{miss}}$ distributions in the low-pileup and high-pileup regime, for the true and generated events.\label{fig:comparisonMETvsNvtx}}
\end{figure*}

Further insight can be derived comparing the correlation matrix between all variables,  shown in Fig.~\ref{fig:comparisonCorrelation} for both the target distribution and the trained model. On a qualitative level, the GAN model is capable of learning the main structures observed in the original dataset, but the quantitative agreement is poor. Despite this overall lack of precision, the GAN is capable of learning some non trivial correlation between variables. For instance, 
the left plot of Fig.~\ref{fig:comparisonMETvsNvtx} shows that the $Z$ boson peak is fairly (but not perfectly) modeled. The right plot in the same figure shows that $\MET$ distribution is correctly learned both for large and small pileup events, i.e., for events with $>25$ and $<15$ vertices, respectively. 

\begin{figure*}[htb!]
\centering
\includegraphics[width=\textwidth]{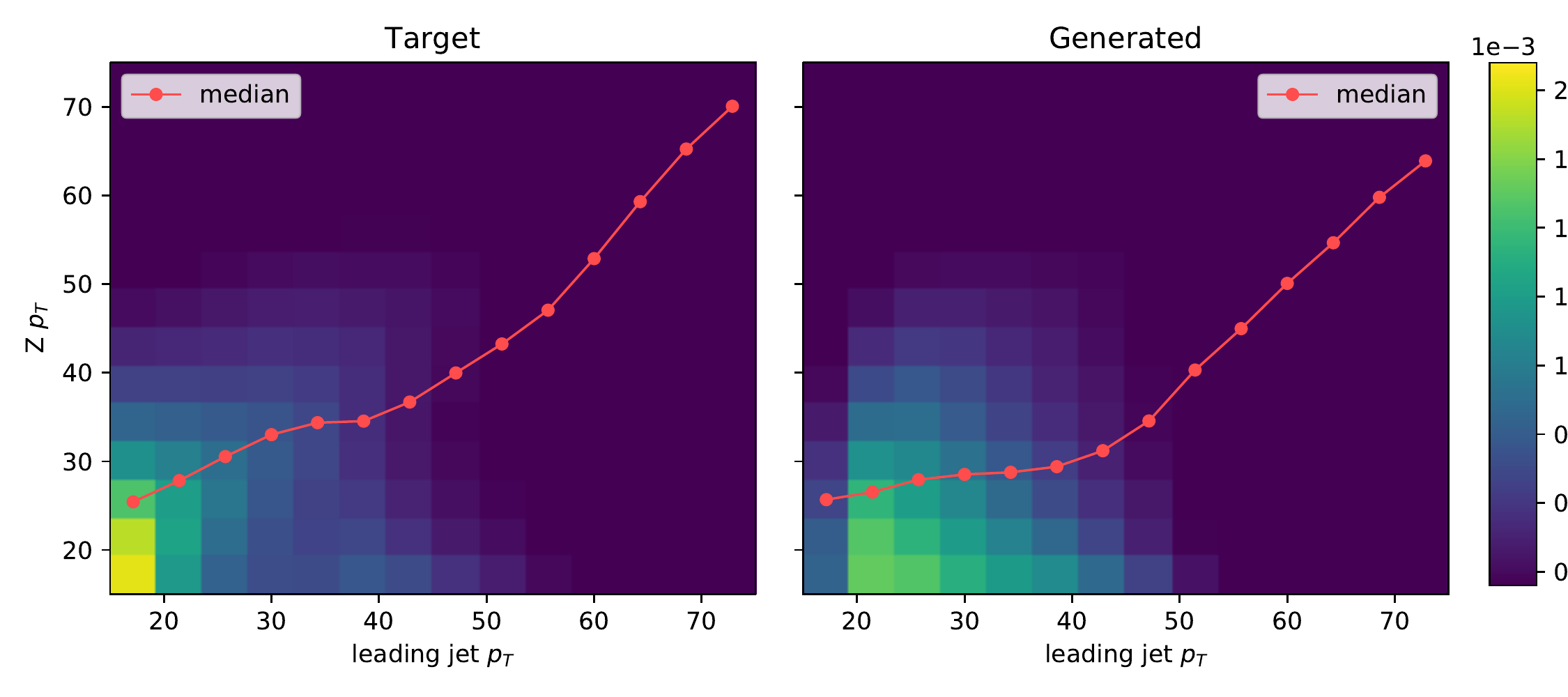}
\caption{Comparison of Z-boson transverse momentum plotted against the leading-jet $\pt$, for target (left) and generated (right) samples.\label{fig:comparisonZptvsJetpt}}
\end{figure*}

Finally, Fig.~\ref{fig:comparisonZptvsJetpt} shows the correlation between $\pt^Z$ and the leading-jet $\pt$, which is a direct consequence of $\pt$ conservation . The correlation is expected to be roughly linear ($\pt^Z \propto \MET$), since the $Z$ boson primarily recoils against the highest-\pt jet. 
This relation is observed for $p_T>40$~GeV and correctly modelled by the GAN, as shown by the relation between the median $Z~\pt$ and the jet \pt, represented by the orange line
in both plots. 

\section{Remarks on performance}
\label{sec:remarks}

Results in Fig.~\ref{fig:results_minigan_bce} show that GANs can learn the multi-dimensional pdf of $\cal{O}$(10) features. In addition, the cost function in Eq.~(\ref{eq:gen_loss}) provides a performance enhancement by incorporating regression terms into the loss function. The usefulness of this method is boosted due to the empirical observation that models which performed best for a small subset of features tended to perform better in other variables as well. Figures~\ref{fig:results_fullgan}~and~\ref{fig:comparisonCorrelation}  indicate that a fair performance is still possible with a larger number of quantities but the reached precision is still insufficient to meet the precision requirements of an LHC data analysis. The GAN shows problems in learning distributions with sharp features such as edges, spikes (see for instance jet $\pt$'s and lepton isolation quantities) and multiple peaks. This inaccuracy is then propagated to some of the quantities computed from the generated features, e.g. $\mll$ and $\metphi$. One could cure this problem  training separate subsets of the data (e.g., exclusive jet multiplicities), or trading sharply distributed quantities for some smoother function of them (e.g., $\log\pt$ rather than $\pt$). As an example of this, we consider the case of discrete quantities, for which we find it convenient to apply a Gaussian smearing.
This approach could be used in general for any discrete quantity, but it is also true that a use-case-dependent clever design of the dataset might avoid the need to deal with some of the discrete features (see Appendix~\ref{sec:discrete_quantities}). For instance, when dealing with opposite-charge lepton pairs, one could order the two leptons by their electric charge (rather than by the $p_T$) and implicitly encode any correlation between the charge and kinematic features into differences in the distribution of the kinematic quantities (e.g., $p_T^{(+)}$~vs.$p_T^{(-)}$). We didn't follow this strategy here, in the spirit of showing the drawbacks as well as the strong points of the approach we propose.

As a final remark, an overall improvement on derived quantities can be achieved adding extra regression terms to the loss function, in the same spirit of what is done in Eq.~(\ref{eq:gen_loss}).


\section{Conclusions and Outlook}
\label{sec:conclusions}

We investigated the possibility of using Generative Adversarial Networks as a tool to create analysis-specific datasets for LHC data analyses. Starting from the consideration that a generic LHC data analysis only uses ${\cal O}(10)$ physics quantities, and using the promising performances of GANs to learn complicated probability distribution functions, we train a set of GAN models to generate new examples. 
We show a fully working example, which highlights the potential of this method, as well as a large-dimension extension where problems with specific features in the dataset are observed. We discuss how these problems are related to discrete features, features exhibiting edges and spikes, as well as multiple peaks. We comment on how the solution to this problem is intrinsically use-case specific and can be handled with specific workarounds. We show how, in general, better performances are obtained when the GAN loss function is modified with regression terms, a fact that was already stressed for image-related GAN models~\cite{JetGAN,CaloGAN1,CaloGAN2}. We also propose an objective criterion, based on the KS test, to select the best model in a quantitative way, i.e., not just relying on a qualitative by-eye assessment of the generator performance. 

With additional consolidation and development, this kind of generative model might be used to increase the statistics of centrally-produced Monte Carlo datasets. As the HEP community faces ever-increasing demand for simulated events, specifically in the era of the High-Luminosity LHC, this possibility could imply a substantial paradigm shift. In this new paradigm, large-scale production of simulated events would be biased toward ``tail distributions" that populate low probability regions of phase space, and for which large equivalent immensities correspond to small number of events to be generated. Generative models could instead be used to augment statistics for the largest samples, reducing the need for full simulation. The biggest upside of this method would be that datasets derived in this manner will be analysis-specific. One one hand, this would reduce the amount of MC to be centrally produced, also allowing to reduce the intermediate storage utilization. On the other, each analysis will require a dedicated training and a dedicated output dataset, with a consequent demand for a centralized network-training facility. We stress the fact that the GPU utilization to train the models for this project was relatively modest and that the emergence of this new workflow is perfectly in line with on-going R\&D projects on train-on-demand solutions in HEP (see for instance Refs.~\cite{mpi-learn,tfaas}).

Io our opinion, analysis-specific GAN models are an interesting research direction to possibly solve the problem of large MC simulation needs, and possibly an answer to the foreseen large MC demand for the High-Luminosity LHC upgrade. Generating the target dataset with {\tt PYTHIA} and {\tt Delphes} was found to be $\sim 3600$ times slower; the generator model is stored in a file which is smaller than 10 MBs, whereas the generated events take up 2 GBs, a reduction of two orders of magnitude in size. 

In addition to the application promoted in this paper, the unsupervised nature of GANs make this kind of method extremely general for generation of discrete events in fields where detailed knowledge of the underlying mathematical dynamics is not available. We hope that this work will motivate further studies to consolidate this strategy into an application-specific set of models.

As a final remark, we stress the fact that a minor strategy modification would allow to extend this work to the problems of fast simulation of detector responses as well as to unfolding. These extensions will be discussed in a future publication

{\bf NOTE ADDED:} while finalizing this paper, Ref.~\cite{Otten:2019hhl} appeared. There, a similar strategy is presented and similar results are obtained. In addition, Ref.~\cite{Otten:2019hhl} provides an interesting comparison between GANs and variational autoencoders. 

\section*{Acknowledgments}
We thank G.~R.~Khattak, S.~Vallecorsa,  P.~Musella, F.~Pandolfi,  P.~Silva, and J.~R.~Vlimant for useful discussions. This project was supported by a grant from the Swiss National Supercomputing Center (CSCS) under project ID cn01. MP  received support from the European Research Council (ERC) under the European Union's Horizon 2020 research and innovation program (grant agreement n$^o$ 772369).

\section*{Appendix A: Discrete Quantities} 
\label{sec:discrete_quantities}

Training GAN models on datasets which included integer quantities, e.g. the number of vertices, was associated with lower performance. This could perhaps have been anticipated, given that discrete valued pdfs are not differentiable on the real line and so can lead to issues with back-propagation. However, our dataset, and presumably many other real datasets to which discrete event generation based on GANs might be applied, will contain a mix of both continuous and discrete or categorical data. Therefore, generic applications of high dimensional GAN-based fits will often need methods for dealing with discrete quantities. We list below several ideas for dealing with discrete quantities, which we experimented in this study:

\begin{itemize}
    \item Quantities like number of vertices, \nvtx, can be considered `psuedo-continuous' in that a minimal change of $\pm 1$ in the count is reflected by a small change in the rest variables. More rigorously, a minimal change in the discrete quantity generates a minimal change in the pdf on the sub-manifold populated by the rest of the quantities. In other words, the distribution of any function of the subset of input vectors with $\nvtx = 25$, should be very close to the distribution of that function with $\nvtx = 24$ or $26$. In this case, a Gaussian smearing (see Section~\ref{sec:data}) is a straightforward procedure to turn the discrete distribution into an analytic one. Taking the floor of the generator output will then return integers which should be distributed very closely to the original discrete distribution.
    
    \item Counts, like the number of jets in an event, can be treated by generating continuous values for each instance and using a cutoff threshold for counting. For instance, to get the number of jets in an event, our event vector contains slots for momenta of the 5 leading energy jets. If less than 5 jets were generated in the event, the additional jet energies are set to 0. In order to construct the number of jets from a generated event, we simply count the number of jet $\pt$s which are greater than 15 GeV.
    
    \item Ordering can be used to make certain categorical distinctions in cases where the number of categories is small or ``dense''. For instance, if the charge of the muons needed to be predicted in our dataset, an easy way to assign charge would be to order the lepton four-momenta by charge rather than \pt.
    
\end{itemize} 
In general, application-specific solutions might be employed, depending upon the nature of the problem.

\medskip

\bibliographystyle{unsrt}
\bibliography{bib}

\end{document}